# TrAD: Traffic Adaptive Data Dissemination Protocol for Both Urban and Highway VANETs

Bin Tian, Kun Mean Hou, Jianjin Li
LIMOS Laboratory CNRS UMR 6158
Blaise Pascal University
Clermont-Ferrand, France
{ bin.tian, kun-mean.hou, jianjin.li }@isima.fr

*Abstract*—Vehicular Ad hoc Networks (VANETs) aim to improve transportation activities that include traffic safety, transport efficiency and even infotainment on the wheels, in which a great number of traffic event-driven messages are needed to disseminate in a region of interest timely. However, due to the nature of VANETs, highly dynamic mobility and frequent disconnection, data dissemination faces great challenges. Inter-Vehicle Communication (IVC) protocols are the key technology to mitigate this issue. Therefore, we propose an infrastructure-less Traffic Adaptive data Dissemination (TrAD) protocol that considers road traffic and network traffic status for both highway and urban scenarios. TrAD is flexible to fit the irregular road topology and owns double broadcast suppression techniques. Three state-of-the-art IVC protocols have been compared with TrAD by means of realistic simulations. The performance of all protocols is quantitatively evaluated with different real city maps and traffic routes. Finally, TrAD gets an outstanding overall performance in terms of several metrics, even though under the worse condition of GPS drift.

## I. INTRODUCTION

Vehicular Ad hoc Networks (VANETs) are believed to be the crucial technology for Intelligent Transportation Systems (ITS) and Traffic Information Systems (TIS), which can be used to support the development of various applications related to traffic safety, transport efficiency and even entertainment on the wheels [1]. However, the main characteristics of VANETs are spatially and temporally localized, highly dynamic and data-intensive, which causes many difficulties to the design of IVC protocols that is the key technology of VANETs [2]. To cope with these issues, the broadcast communication paradigm is usually used to disseminate data messages [3]–[6], since broadcasting is flexible to transmit particular data to a number of uninformed vehicles in a region of interest (ROI). Moreover, several broadcast protocols have been proposed to prove the possibility and the applicability of disseminating data in both urban and highway scenarios [5], [6].

However, there are many challenges to design broadcast protocols. The road traffic environments in urban and highway scenarios are totally different. In urban scenario, there are various road topologies and many buildings beside roads. Several dense networks would be composed around some traffic hubs during rush hour. Due to the existence of vehicular mobility and building obstacles, these dense vehicular networks are partitioned, so that the disconnected network problem is ubiquitous. Therefore, how to suppress the broadcast storm in dense networks and how to fill the gap between these disconnected networks are the crucial challenges. Although the road topology is relatively simple in highway scenario, the high vehicular density and the high speed are also issues to the design of protocols. In addition, IEEE 802.11p standard does not establish a Basic Service Set (BSS) and its Carrier Sense Multiple Access with Collision Avoidance (CSMA/CA) mechanism has not the acknowledgments process. These amendments indeed bring a fast connection process, but leads to an unreliable connectivity of broadcasting.

In this paper, we attempt to address these issues by proposing an infrastructure-less Traffic Adaptive data Dissemination (TrAD) protocol to work seamlessly in both urban and highway scenarios. The TrAD protocol includes two components: one is a broadcast suppression technique that uses a time slot scheme to constrain the broadcast storm problem and improves the reliability of transmission. The other is a store-carry-forward mechanism that not only selects appropriate vehicles to fill the connectivity gap between different disconnected networks, but also *further* suppresses the redundant transmissions. Furthermore, a comprehensive performance evaluation is performed by means of realistic simulations. We compare TrAD with three state-of-the-art IVC protocols in both urban and highway scenarios.

The remainder of this paper is organized as follows. In the next section, we summarize the progress of related literature on this subject, and then present the challenges that need to be resolved. That motivates us to propose the TrAD protocol in section Ⅲ, where every mechanism of TrAD is elaborated. In section Ⅳ, we evaluate the performance of TrAD by means of realistic simulations. Finally, we conclude this paper and present the future work.

## II. RELATED WORK

Most of data dissemination protocols dedicated their contributions to resolve two main problems. One is the broadcast storm problem in the well-connected network. The broadcast suppression technique is used to mitigate this problem. The other is the disconnected network or network partition problem. For this issue, various store-carry-forward mechanisms have been designed.

*A. Broadcast Suppression Technique*

Tseng et al. [7] discovered the broadcast storm problem and proposed three pioneer schemes to alleviate it. Nevertheless, these schemes are specified for Mobile Ad hoc Networks (MANETs). Wisitpongphan et al. [8] proposed some interesting techniques for highway VANETs that include *weighted p-persistence*, *slotted 1-persistence*, and *slotted p-persistence*. Schwartz et al. [9] improved the slotted 1-persistence for their simple and robust dissemination (SRD) protocol. The optimized slotted 1-persistence defines different priorities in two directions of highway. However, all above-mentioned efforts focused on the one-dimensional highway scenario.

In order to handle broadcast storm in two-dimensional urban scenario, eMDR uses real roadmap and GPS information to identify the location of vehicles, so that it can operate the data dissemination [10]. Viriyasitavat et al. [4] proposed the UV-CAST protocol where an illustrative equation is presented to calculate a wait time to rebroadcast in terms of position and distance. It assigns a shorter rebroadcast delay to the vehicle located at an intersection, which intends to disseminate data message in more directions. However, Fogue et al. [11] pointed out that eMDR and UV-CAST should not blindly trust the GPS positioning that is usually not accurate. Thus, the impact of GPS drift on the performance of protocols should be evaluated.

DRIVE and AMD protocols provided solutions to classify neighbors into different quadrants or sectors, which can support to disseminate data in both urban and highway scenarios [5], [6]. DRIVE divides the communication area into four quadrants and selects a sub-area in each quadrant called *sweet spot*. The vehicles in sweet spot get a shorter rebroadcast delay. AMD adaptively separates the vehicle's communication area into 2 or 4 equal sectors according to the road topology and the distribution of neighbors. However, the road topology in the real world is not always regular like Manhattan-Grid style. The approaches of these two protocols are not flexible enough to fit irregular road topologies.

*B. Store-Carry-Forward Mechanism,*

For disconnected network problem, most of the literature mainly focused on the selection of *SCF-agent* vehicles that can store and carry the data message until they meet new opportunities to forward. UV-CAST attempts to select the vehicles located at the boundary of connected network to be SCF-agents by a gift-wrapping algorithm. Since this algorithm is a distributed version, the set of SCF-agent vehicles is always a superset of all boundary vehicles [4]. Thus, UV-CAST would trigger more redundant transmissions. DRIVE does not resort to beaconing, where it only sets a timer for the vehicles outside the ROI to schedule the rebroadcast [6]. It is obvious that most opportunities of transmissions would be missed. AMD does not suffer this issue, it uses beaconing that can trigger a new broadcast when a vehicle make transition from a role of *tail* to *non-tail* in one of its directional sectors [5]. Nevertheless, the directional sector classification of AMD can not accurately recognize the irregular road topology, which would lead to a failure to detect some transitions so as to miss some opportunities to forward data messages.

III. TRAFFIC ADAPTIVE DATA DISSEMINATION

*A. Concepts*

Several concepts of TrAD are defined, which will be used throughout the paper. The illustration of TrAD protocol is shown in Fig.1.

- *Directional Cluster $C_d$*: It is a group of vehicles in the neighborhood of a sender *S*, which are in a similar direction with respect to the sender *S*.

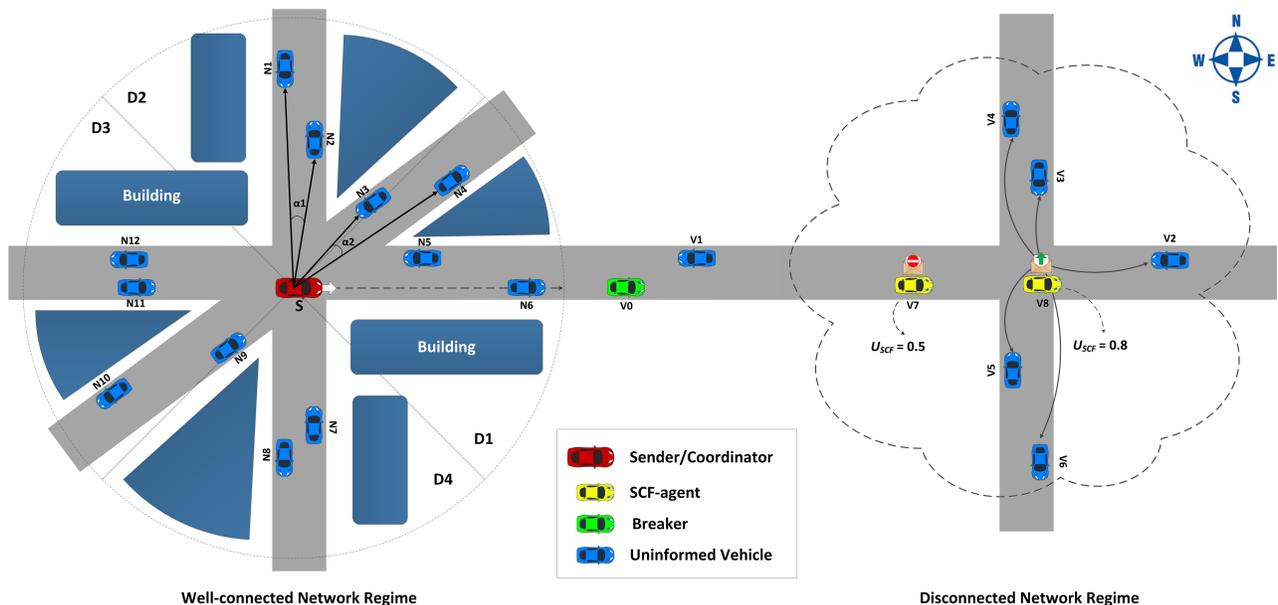

Fig.1. Illustration of TrAD protocol.

- *Coordinator*: This concept has been proposed in GPCR protocol [12]. The coordinator is the vehicle that is located at an intersection. For instance, the sender *S* in Fig.1. (the red vehicle) is a coordinator.
- Breaker: In a well-connected network, the *breaker* is not only the furthest vehicle but also the one which moves towards outside of the network. For instance, the green vehicle V0 in Fig.1 is a breaker.

*B. The Protocol*

In this section, we propose the principle of operation for the TrAD protocol, which consists of two main components: the broadcast suppression technique for well-connected network and the store-carry-forward mechanism for disconnected network (Fig.2). TrAD requires beaconing to maintain the up-to-date status in a one-hop neighborhood. The structure of beacon is < *Beacon ID, Sender ID, Global GPS Position, Driving Direction, Number of Neighbors, Channel Busy Ratio, Message List* >. The message list entry contributes to recognize the uninformed vehicle by comparing neighbor's message list with local message list.

*1) Broadcast Suppression Technique*

The protocol makes a centralized decision in the sender to control the rebroadcast order of neighbors, since the centralized decision can resolve the hidden terminal problem in the multi-directional dissemination.

*a) Vector-angle-based cluster classification mechanism*

This classification mechanism only resorts to position information without road map, so that it can fit more complex road topologies. The algorithm uses vector angle to identify whether the vehicles belong to a directional cluster.

The operation is described as follows (Fig.1): In the well-connected network, the sender *S* extracts the first neighbor $N1$ from its neighbor list and classifies $N1$ into the directional cluster $C_1$. Then, the algorithm calculates the vector $\vec{v_1}$ from *S* to $N1$ ($\vec{v_1} = \overrightarrow{SN1}$). After that, *S* extracts next neighbor $N2$ and forms another vector $\vec{v_2}$ from *S* to $N2$. The angle $\alpha_1$ between $\vec{v_1}$ and $\vec{v_2}$ can be obtained from their dot (or scalar) product ($\alpha_1 \in [0, \pi]$). It is compared with a threshold angle $\alpha$ ($\alpha = 10°$). If the angle $\alpha_1$ is less than $\alpha$, the neighbor $N2$ is classified into the directional cluster $C_1$. If not, neighbor $N2$ still remains in the neighbor list for next step. In Fig.1, $\alpha_1 < \alpha$, so $N2 \in C_1$. This process continues until all neighbors have been checked. Therefore, we can identify a group of vehicles $C_1$ that are located in similar directions as $N1$. The members of $C_1$ are saved in a cluster buffer and eliminated from next classification step. The classification process continues to be performed until all neighbors have been classified into respective directional clusters.

The classification result of TrAD in Fig.1 is $C_1=\{N1, N2\}$; $C_2=\{N3, N4\}$; $C_3=\{N5, N6\}$; $C_4=\{N7, N8\}$; $C_5=\{N9, N10\}$; $C_6=\{N11, N12\}$. Vehicles in the same direction with respect to *S* are exactly classified into the same directional cluster. Additionally, the classification result of AMD is also illustrated. AMD divides the radio range into 4 equal sectors (D1, D2, D3 and D4). However, the classification result is not accurate. The $C_2=\{N3, N4\}$ and $C_3=\{N5, N6\}$ are merged into sector D1, and the $C_5=\{N9, N10\}$ and $C_6=\{N11, N12\}$ are combined into sector D3. Moreover, vehicles *N3* and *N9* are ambiguous on the boundary of sectors. The result proves that the directional sector classification of AMD is inflexible and inadaptable for irregular road topology.

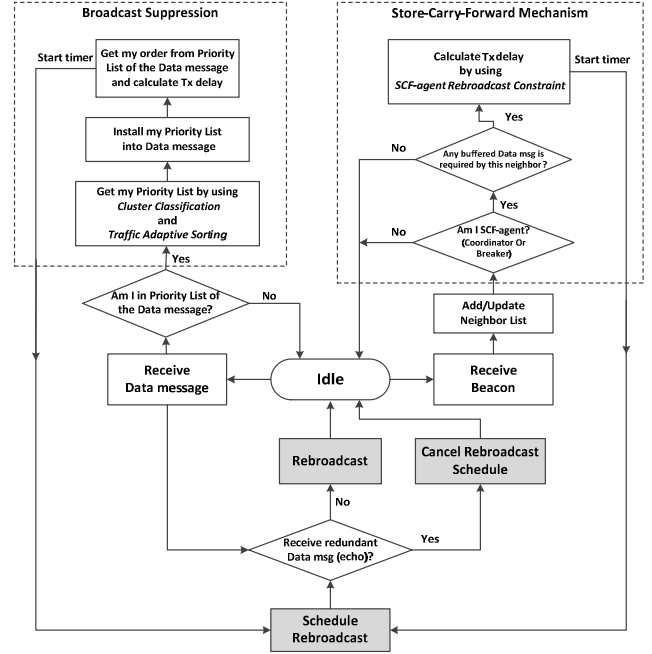

Fig.2. The operation of TrAD protocol flowchart

*b) Traffic adaptive sorting technique*

This technique considers both road traffic and network traffic status, which includes three components: the number of neighbors *N* and the distance between sender and neighbor *D* represent the road traffic condition, and the Channel Busy Ratio *CBR* reflects the network traffic status.

The metric of the number of neighbors *N* (Eq.1) indicates the coverage of potential vehicles in the next transmission. #Neighbor is the instantaneous number of neighbors in neighbor list. max.#Neighbor is a preset maximum number of neighbors.

$$N = \min\left\{\frac{\#Neighbor}{max.\#Neighbor}, 1\right\} \quad (1)$$

*D* is the metric of the distance between a sender *S* and a neighbor (Eq.2). $Dist_{sn}$ is the distance between a sender *S* and a neighbor. $max.RadioRange$ is the maximum communication range of the wireless access medium.

$$D = \min\left\{\frac{Dist_{sn}}{max.RadioRange}, 1\right\} \quad (2)$$

The metric *CBR* (Eq.3) describes the channel usage of local node in a time interval (1s). The time usage of busy channel $T_{busy}$ is got from PHY layer of IEEE 802.11p.

$$CBR = \frac{T_{busy}}{Time\ Interval} \quad (3)$$

Thereby, we propose an equation (Eq.4) to assign a higher transmission utility $U_{TX}$ to the neighbor that owns more neighbors $N$, further distance $D$ and smaller $CBR$.

$$U_{TX} = (1 + N) \cdot (1 + D) \cdot (2 - CBR) \quad (4)$$

According to Eq.4, we get the $U_{TX} \in [1,8]$ of every neighbor. First, we sort the neighbors in every directional cluster in terms of their $U_{TX}$ in decreasing order. Then, a *round-robin* fashion is used to define the final order of neighbors in *priority list*, where the vehicle with the highest $U_{TX}$ in the directional cluster $C_1$ is put in the first position, followed by the vehicle with highest $U_{TX}$ in directional cluster $C_2$, and so forth [5]. This fashion guarantees the fairness for each directional cluster.

The sender $S$ inserts the priority list into the header of data message before transmission. The format of header is < *Data ID, Originator ID, Sender ID, Originator Position, Sender Position, Priority List* >. When the recipients receive the data message from $S$, they extract the priority list and find their order $R \in [0, n-1]$, n is the total number of neighbors. TrAD allocates one slot time $st$ for each vehicle. The transmission delay $Delay_{TX}$ is computed by using Eq.5.

$$Delay_{TX} = st \cdot R \quad (5)$$

*2) Store-Carry-Forward (SCF) Mechanism*

SCF mechanism includes two components: one is responsible for selecting appropriate vehicles to be SCF-agent. The other is a special operation to *further suppress* the redundant broadcasts.

*a) The Selection of SCF-agent*

Here, we introduce the techniques to identify the coordinator and breaker. A vehicle that satisfies any one condition of them is selected to be a SCF-agent.

- *Coordinator:* Since an intersection position list is preloaded into the system, every vehicle can check the distance between itself and all intersections when the vehicle receives a beacon message. If any distance is less than 20m, the vehicle is believed to be a coordinator.

- *Breaker:* When the vehicle receives a data message, the protocol will check and eliminate the possibility of coordinator role. After that, the vehicle checks whether its driving direction is the same as data forwarding direction. If so, the vehicle will search whether there is a further neighbor driving in the data forwarding direction. If not, the vehicle is defined as a breaker. It is worth mentioning that this procedure will iterate until the boundary of well-connected network.

The coordinator is specified for the two-dimensional topology of urban scenario. The heading of coordinator after passing an intersection is various and can not be predicted. Therefore, the coordinator is beneficial to discover more uninformed vehicles. The breaker is defined for both urban and highway scenarios. It can carry the data message and drive towards outside of well-connected network.

*b) SCF-agent Constrained Rebroadcast Technique*

This technique aims to trigger the rebroadcast of SCF-agent and constrain it simultaneously, if multiple SCF-agents receive a same request by beaconing. Thus, we design a distributed time slot scheme to calculate different broadcast delay for every SCF-agent. If a SCF-agent receives a redundant data message (echo) that is the same as the one it is scheduling, the SCF-agent will cancel the schedule and switch to an idle state.

Normally, the uninformed vehicles drive toward the SCF-agents from outside of the connected network where the SCF-agents get the data message. Therefore, we design the Eq.6 to calculate a utility of a SCF-agent $U_{SCF}$ ($U_{SCF} \in [1,4]$), which assign a higher $U_{SCF}$ to the SCF-agent owns closer distance $D$ to the uninformed vehicle and smaller channel busy ratio $CBR$.

$$U_{SCF} = (2 - D) \cdot (2 - CBR) \quad (6)$$

A broadcast delay $Delay_{SCF}$ for a SCF-agent is calculated by Eq.7. This equation assigns a shorter delay to the SCF-agent with a higher $U_{SCF}$. Notice that a new incoming uninformed vehicle normally does not only require one data message. It is possible to require several data messages once. If the protocol does not control this behavior properly, a burst of transmissions will be incurred. Therefore, we propose the Eq.7 to calculate the $Delay_{SCF}$:

$$Delay_{SCF} = st \cdot \left[BurstCount + \left(1 - \frac{U_{SCF}}{4}\right)\right] \quad (7)$$

We give a basic slot time $st$ between each required data message. The $BurstCount$ is the transmission order of data messages. For the same data message, the $st \cdot \left(1 - \frac{U_{SCF}}{4}\right)$ part lets a SCF-agent with a higher utility $U_{SCF}$ transmit first. This part also works as an additional delay, which can mitigate the time slot boundary synchronization problem.

For instance, in Fig.1, vehicles V7 and V8 are the SCF-agents and carry new data messages. They received the request from the vehicles around them. Since the utility $U_{SCF}$ of vehicle V8 (0.8) is larger than the one of V7 (0.5), vehicle V8 is triggered to broadcast data messages first. V7 hears the redundant data and cancels the schedule of the same data message. Thus, the redundant transmissions are suppressed.

IV. PERFORMANCE EVALUATION

The performance of TrAD protocol is evaluated by means of realistic simulations in both urban and highway scenarios. A *scenario* includes a map and a series of traffic routes. The *map* is composed of road topology and building obstacles, and the *traffic route* specifies the planned round-trip route for every vehicle cluster. Several traffic routes comprise the road traffic network on the map. We set the departure of vehicles following an exponential distribution. The programs are implemented to simulate three state-of-the-art IVC protocols in order to compare with TrAD protocol, namely DV-CAST for highway scenario [3], UV-CAST for urban scenario [4] and AMD for both urban and highway scenarios [5].

We use the Veins 3.0 framework[1] that is based on two simulators, OMNeT++ 4.4.1, an event based network simulator[2], and SUMO 0.23.0, a road traffic simulator[3]. IEEE 802.11p standard has been used to be the MAC and PHY layer for DSRC/WAVE. The data rate is set to 6 Mbit/s that is the default data rate of broadcasting in IEEE 802.11p. The transmission power is set to 300 mW. The Friis Free Space Path Loss (FSPL) propagation model is used, where the exponent α is assigned to 3.0, as it is in the range [2.7, 5.0] that is estimated for outdoor shadowed urban environment [13]. The radio range reaches approximately 366m according to the setting of propagation model. The Bit Error Rate (BER) model provided in Veins is proposed by [14]. The shadowing obstacle model treats the buildings (pink polygon in Fig.3) as obstacles [15]. For all simulations, the source node broadcasts the data message every 2s. The beacon message of TrAD is sent every 1s and its lifetime is 1.5s. The size of data message is 2312 bytes. The size of beacon message is 378 bytes, in which the entry number of message list is 40. The vehicle density is from 40 v/km$^2$ to 160 v/km$^2$, the interval is 20 v/km$^2$. The setting of simulation system is shown in TableⅠ.

TABLE Ⅰ SIMULATION SETTING

| Physical layer | Frequency band | 5.89 GHz |
|---|---|---|
| | Bandwidth | 10 MHz |
| | Tx power | 300 mW |
| | Receiver sensitivity | -100 dBm |
| | FSPL exponent α | 3.0 |
| | Thermal noise | -110 dBm |
| | Radio range | ~366m |
| Link layer | Bit rate | 6 Mbit/s |
| | CW | [15,1023] |
| | Slot time | 13 μs |
| | SIFS | 32 μs |
| Data Broadcasting | Broadcast frequency | 0.5 Hz |
| | Data size | 2312 bytes |
| Beaconing | Beacon frequency | 1 Hz |
| | Beacon size | 378 bytes |
| | Message list entries | 40 |
| TrAD | $st$ | 8ms |
| | $\alpha$ | 10° |
| | $max.\#Neighbor$ | 20 |
| | $max.RadioRange$ | 366m |
| AMD | $st$ | 5ms |
| | $ts_d$ | 1 |
| | $AD_{ij}$ | DIFS |
| UV-CAST | $\tau_{max}$ | 500ms |
| DV-CAST | $st$ | 5ms |
| | $N_{st}$ | 5 |
| | WAIT Ⅰ | 120s |
| | WAIT Ⅱ | 120s |

We created one highway scenario and two urban scenarios. Geographic data were retrieved from the OpenStreetMap[4] (OSM) database. The traffic rule and traffic light were set according to the real world ones. For urban scenario, two maps are used: one is a fragment of Manhattan borough of

---

1. Veins: http://veins.car2x.org/
2. OMNeT++ : https://omnetpp.org/
3. SUMO: http://sumo.sourceforge.net/
4. OpenStreetMap: www.openstreetmap.org

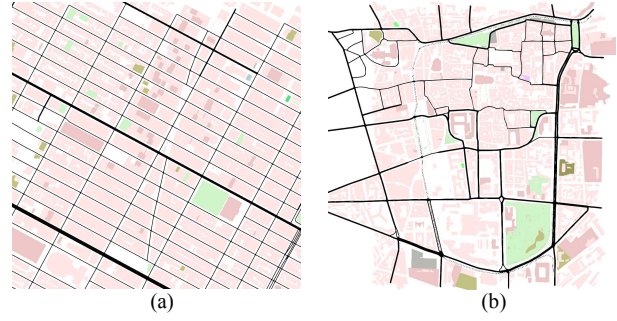

Fig.3. Maps: (a) Manhattan borough of New York City (USA). (b) Clermont-Ferrand (France)

New York City, USA (Fig.3a), the other is the downtown area of Clermont-Ferrand, France (Fig.3b). The size of both maps is 1 x 1 km$^2$. The data message is generated by a fixed node (source) located at an intersection in the center of each map and collected by all mobile vehicles in the ROI. The highway map is selected from a section of A711 highway near Clermont-Ferrand airport (Aulnat), France. It consists in 2km straight long and two lanes in each direction. The data message is created by one fixed node (source) at the west end of the highway and gathered by another fixed node (receiver) at the east end.

### A. Maps

The statistics of maps (TableⅡ) shows that the map of Clermont-Ferrand has more lanes, junctions and shorter street lengths than Manhattan, which means that the traffic environment is more complex in Clermont-Ferrand than in Manhattan. Similar traffic routes are created for both maps, which lets vehicles be distributed uniformly in the scenario.

TABLE Ⅱ STATISTICS OF MAPS

| City Map | Clermont-Ferrand | Manhattan |
|---|---|---|
| Total lanes | 366 | 166 |
| Total junctions | 137 | 86 |
| Avg. street length | 97.39 | 151.45 |
| Avg. lanes/street | 1.62 | 1.11 |

According to the Fig.4 and Fig.5, TrAD uses moderate number of transmissions (overhead) to achieve better or similar packet delivery ratio (PDR) and delay compared with UV-CAST. Although UV-CAST gets similar PDR or delay with TrAD in a few experiments, it consumes too much overhead. In contrast, AMD well controls its overhead. However, it fails to reach qualified PDR and delay compared with other two protocols. Therefore, TrAD has an outstanding performance cost ratio.

Specifically, the complex map of Clermont-Ferrand constrains the transmissions of TrAD. In consequence, the PDR of TrAD in Clermont-Ferrand fails to achieve 90% when the vehicular density is less than 80 v/km$^2$. By contrast, the PDR of TrAD in Manhattan is more than 90% at all the vehicular densities. But when the vehicular density is more than 80 v/km$^2$ (include 80 v/km$^2$), the PDR of TrAD in Clermont-Ferrand exceeds 90% and the usage of overhead is

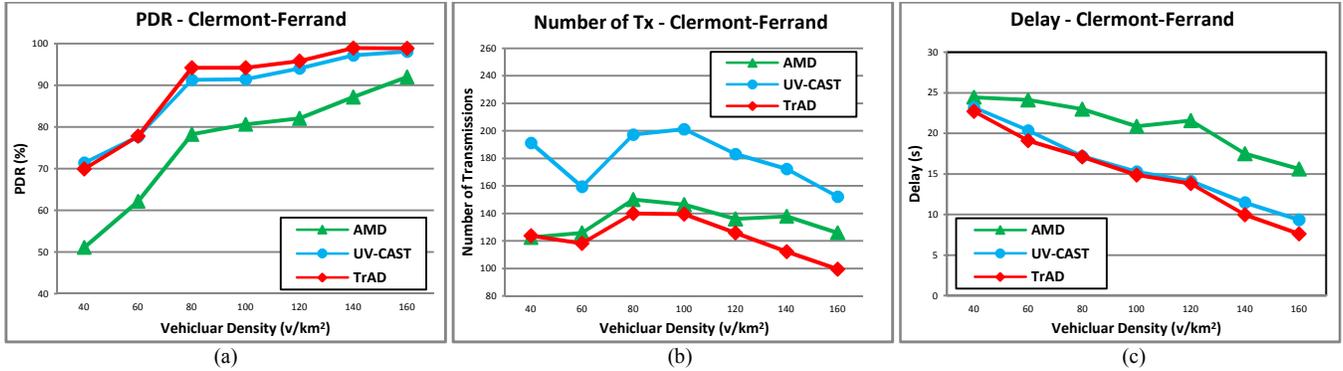
Fig.4. Result in Clermont-Ferrand scenarios with uniformly distributed traffic route: (a) PDR, (b) Number of Transmissions, (c) Delay.

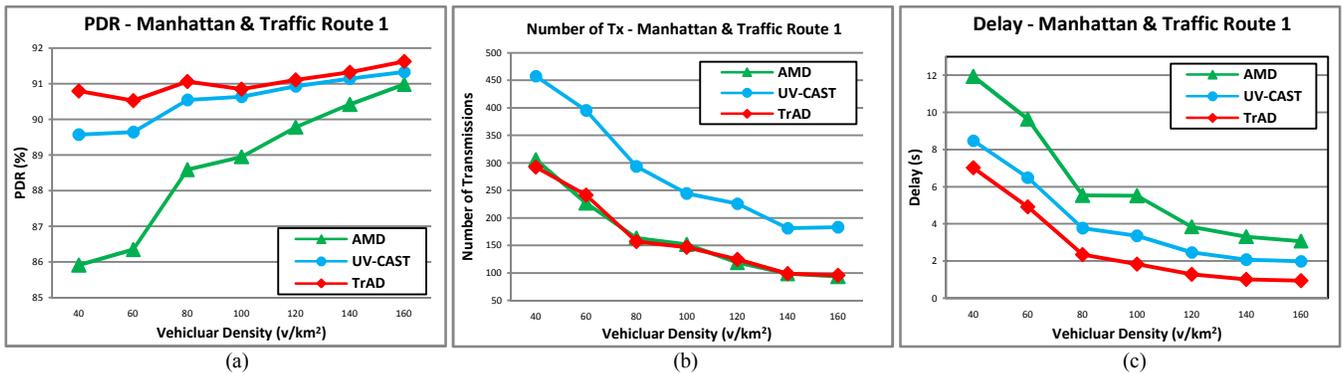
Fig.5. Results in Manhattan scenarios with uniformly distributed traffic route (traffic route 1): (a) PDR, (b) Number of Transmissions, (c) Delay.

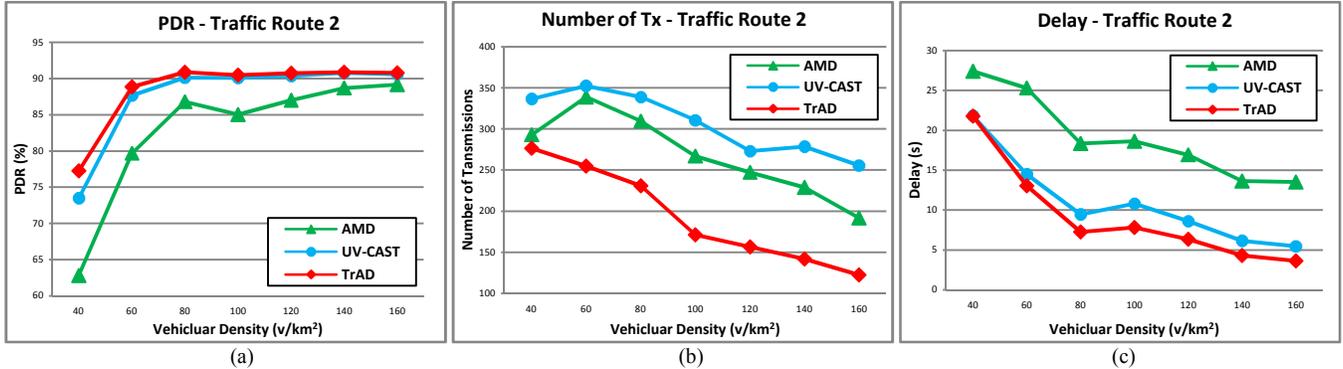
Fig.6. Results in Manhattan scenarios with non-uniformly distributed traffic route (traffic route 2): (a) PDR, (b) Number of Transmissions, (c) Delay.

at roughly the same level as Manhattan. That means TrAD is capable to adapt complex road topology.

### B. Traffic Routes

For urban scenarios, most of previous works used different maps to evaluate the performance of protocols. However, we realize that different traffic routes on a same map could also significantly influence the performance of protocols. Therefore, we designed two different traffic routes based on the Manhattan map (Fig.7). Traffic route 1 (Fig.7a) is uniformly distributed that is *the same traffic route of Manhattan scenario in Fig.3a*. It lets vehicles drive between upper and lower area of the map, so that they can cross the middle line where the source node is located. Thus, the vehicles get more chances to connect to the source node or other vehicles that carry new data. In the opposite, traffic route 2 (Fig.7b) is non-uniformly distributed. It constrains vehicles inside the upper or the lower area of the map. Consequently, only a few vehicles pass by and connect to the source node. Therefore, traffic route 2 could create more disconnected networks than traffic route 1.

As shown in Fig.5 and Fig.6, TrAD still maintains a relatively higher performance cost ratio compared with other two protocols. The gap of number of transmissions between TrAD and AMD is enlarged in the scenario with traffic route 2 compared with the case of traffic route 1. The PDR of TrAD in the scenario with traffic route 2 first reaches 90% and catches up with the same level as traffic route 1 case just after 60 v/km$^2$ vehicular density. This result illumines us that TrAD can mitigate the influence of disconnected network.

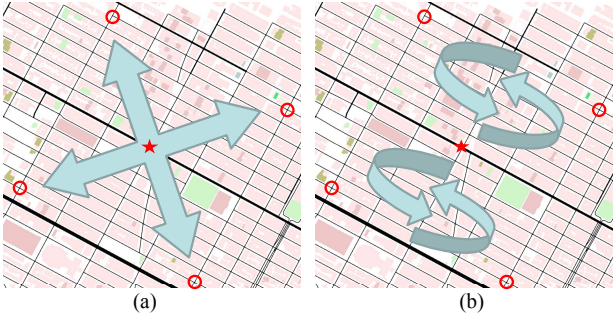

Fig.7. Traffic routes. (a) Traffic route 1. (b) Traffic route 2. The arrows indicated the macroscopical traffic route. The red start is the source node. The four red circles indicate the boundary points of ROI.

In summary, the complex map and non-uniform traffic route indeed impact the performance of data transmission protocols. However, experiment data shows that TrAD costs moderate transmissions to achieve a relatively high PDR and low delay. The reason lies in two factors: one is that TrAD uses the flexible cluster classification to identify the complex road topology, and the traffic adaptive sorting algorithm to select appropriate vehicles to rebroadcast the data message. The other is that the SCF mechanism of TrAD not only discovers more potential uninformed vehicles, but also suppresses the redundant transmissions of SCF-agents. In contrast, UV-CAST triggers more transmission by using its gift-wrapping algorithm and there is no suppression technique in SCF mechanism. The sector classification of AMD is not flexible for the complex map and its SCF mechanism fails to discover more potential uninformed vehicles.

### C. Network Density

#### 1) Urban Scenario

The simulation results of different maps and traffic routes in urban scenarios with the increasing of network density have been elaborated. We realize that the network density, map and traffic route are three main impact factors to the performance of IVC protocol. Each of them can influence the final result. Normally, a lower network density leads to more disconnected networks, which lets protocols get relatively low PDR, high delay and high number of transmissions (overhead). In contract, a higher network density let protocols achieve the opposite results. However, the geographic and traffic environment could also significantly affect the performance of protocols.

#### 2) Highway Scenario

To count the *traffic flow,* inductive loop detectors are deployed under every lane. The unit of traffic flow represents the number of vehicles passed the detection point per hour (vph). We set 5 levels of traffic flow that include 450 vph, 896.4 vph, 1353.6 vph, 1803.6 vph and 2257.2 vph.

Fig.8a shows that TrAD is more scalable to maintain high and stable data coverage. Furthermore, the delay of TrAD is excellent at most experiments and decreases rapidly with the increasing of traffic flow (Fig.8.c). In Fig.8b, TrAD gains the advantage over AMD in sparse networks. While AMD consumes fewer transmissions in dense networks, which conforms with the result of experiments in [7]. However, AMD fails to touch the level of PDR and delay of TrAD. For DV-CAST, its PDR and delay are unstable and worse than others.

SCF mechanism of TrAD and AMD mainly influences their performance in highway scenario. The breaker of TrAD always carries the data and drives towards the receiver. However, AMD could fail to detect opportunities to transmit, so that a part of PDR is lost and the delay is increased. The unreliability of DV-CAST lies in its complex operations that resort to several connectivity flags and timers.

### D. GPS Drift

The proposed protocol is partly based on positioning. However, the GPS system is usually not accurate. Especially in urban scenario, the high-raise buildings could obstruct the signal. Therefore, we investigated the GPS drift in terms of data dissemination speed that represents the percentage of vehicles that are covered by data over time. Five error deviations (0m, 25m, 50m, 75m and 100m) have been injected into the mobility module of simulation, a moderate vehicle density 100 v/km$^2$ is used in the complex map (Clermont-Ferrand). The source node only sends one data message and the data coverage is recorded over time.

The tolerance of TrAD to GPS drift is evaluated (Fig.9). We can observe that the simulation without GPS drift gets the best result. The 25m, 50m and 75m error deviations lightly impact the performance of TrAD, but the data dissemination speed of these deviations still keep close and achieve a relatively middle level result. The result of 100m deviation is worse than others, which can be predicted. But the data coverage of all the deviations reaches 100% at last.

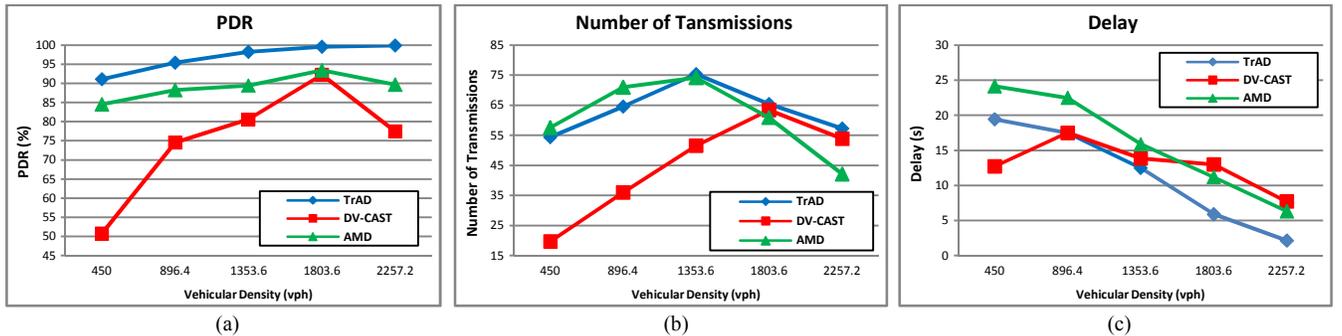

Fig.8. Results in Highway Scenario: (a) PDR, (b) Number of Transmissions, (c) Delay.

TrAD is compared with AMD and UV-CAST in 50m error deviation (Fig.10). TrAD keeps a relatively good upward trend than other two protocols and finally reaches 100% data coverage. However, GPS drift constrains the performance of UV-CAST and lets its dissemination speed slower than AMD at most of time. Finally, UV-CAST and AMD just achieve the data coverage around 80%. These results verify that TrAD is more robust to the impact of GPS drift.

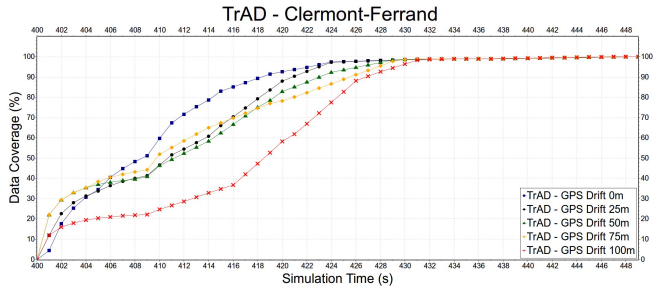

Fig.9. The impact of GPS drift on TrAD in Clermont-Ferrand.

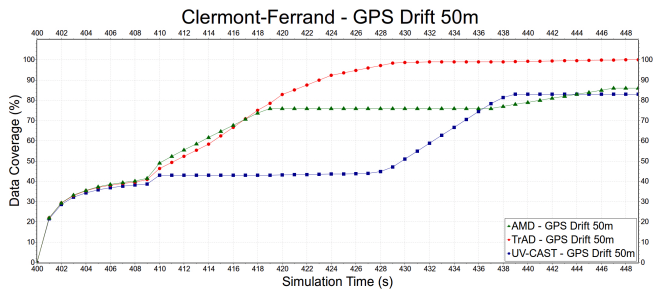

Fig.10. The impact of 50m GPS drift on TrAD, AMD and UV-CAST.

## V. CONCLUSION AND FUTURE WORK

We have proposed the TrAD protocol to improve the performance of data dissemination for both highway and urban scenarios. The outstanding performance is verified by means of realistic simulations, in which three state-of-the-art data dissemination protocols are selected to compare with TrAD. On one hand, TrAD can fit the irregular road topology and considers both road traffic and network traffic status. On the other hand, TrAD uses double suppression mechanisms to limit transmissions, which not only performs the broadcast suppression technique in well-connected network, but also constrains the redundant rebroadcast of SCF-agent in disconnected network. Therefore, TrAD only consumes moderate transmissions to achieve relatively high data coverage with low delay even though suffering GPS drift. For the future work, we plan to implement TrAD protocol on our Wireless Multimedia Sensor Board (MiLive)[5] to perform the real world measurement for specific applications of VANETs.

## ACKNOWLEDGMENT

This work has been sponsored by the French government research program "Investissements d'avenir" through the IMobS3 Laboratory of Excellence (ANR-10-LABX-16-01), by the European Union through the program "Regional competitiveness and employment 2007-2013" (ERDF–Auvergne region), and by the Auvergne region.

Thanks to Philippe Vaslin (Associate Professor, Ph.D.) and Jean Connier (Ph.D. student) for revising the manuscript of this paper. Thanks to Dongsheng Yan (M.E.) and Philip Dechant (M.E.) for creating the realistic scenarios.

---

5. MiLive: http://edss.isima.fr/sites/smir/project/intro?prjId=201